\documentclass{article}
\usepackage{amsmath}

\usepackage{arxiv}

\usepackage[utf8]{inputenc} % allow utf-8 input
\usepackage[T1]{fontenc}    % use 8-bit T1 fonts
\usepackage{hyperref}       % hyperlinks
\usepackage{url}            % simple URL typesetting
\usepackage{booktabs}       % professional-quality tables
\usepackage{amsfonts}       % blackboard math symbols
\usepackage{nicefrac}       % compact symbols for 1/2, etc.
\usepackage{microtype}      % microtypography
\usepackage{lipsum}
\usepackage{graphicx}
\graphicspath{ {./images/} }

\title{Spectroscopic Ellipsometry for Two-Dimensional Materials: Methods, Optical Modeling, and Emerging Phenomena}

\author{
 Ersyzario Edo Yunata \\
  Department of Physics\\
  Faculty of Science and Technology\\
  Airlangga University\\
  Surabaya 60115, Indonesia \\
  \texttt{ersyzario.edo@fst.unair.ac.id} \\
  \And
 Angga Dito Fauzi \\
  Department of Physics\\
  Faculty of Science and Technology\\
  Airlangga University\\
  Surabaya 60115, Indonesia \\
  \texttt{a.dito.fauzi@fst.unair.ac.id} \\
   \And
 Khoirunnisa Qoulan Aziza \\
  Department of Physics\\
  Faculty of Science and Technology\\
  Airlangga University\\
  Surabaya 60115, Indonesia \\
  \texttt{khoirunnisa.qoulan.aziza-2023@fst.unair.ac.id} \\
  \And
 Priscelia Arie Novita \\
  Department of Physics\\
  Faculty of Science and Technology\\
  Airlangga University\\
  Surabaya 60115, Indonesia \\
  \texttt{priscelia.arie.novita-2023@fst.unair.ac.id } \\
  \And
 Dian Meilanita Edi Sayom \\
  Department of Physics\\
  Faculty of Science and Technology\\
  Airlangga University\\
  Surabaya 60115, Indonesia \\
  \texttt{dian.meilanita.edi-2023@fst.unair.ac.id} \\
   \And
Novita Aulia Rafi \\
  Department of Physics\\
  Faculty of Science and Technology\\
  Airlangga University\\
  Surabaya 60115, Indonesia \\
  \texttt{novita.aulia.rafi-2023@fst.unair.ac.id} \\
   \And
 Nabilah Mufidah \\
  Department of Physics\\
  Faculty of Science and Technology\\
  Airlangga University\\
  Surabaya 60115, Indonesia \\
  \texttt{nabilah.mufidah-2023@fst.unair.ac.id} \\
}

\begin{document}
\maketitle
\begin{abstract}
This review provides a comprehensive examination of the role of Spectroscopic Ellipsometry (SE) in elucidating the optical properties of two-dimensional (2D) materials, including graphene, monolayer and multilayer transition-metal dichalcogenides (TMDs), and related organic thin films. By synthesizing experimental methodologies, fundamental ellipsometric principles, optical modeling strategies, and advanced analytical approaches, this work highlights the exceptional capability of SE to resolve dielectric functions, excitonic resonances, optical anisotropy, interlayer interactions, and substrate-induced effects. Recent studies demonstrate SE’s effectiveness in revealing key phenomena such as environment- and synthesis-dependent excitonic shifts, redshifted $\pi \rightarrow \pi^\ast$ transitions in graphene on nickel, high refractive indices in multilayer TMDs, and naturally occurring hyperbolic dispersion in metallic TMDs. Despite its strengths, SE remains constrained by model dependency, limited spatial resolution, and the challenges posed by non-ideal or structurally complex samples. Overall, this review consolidates current knowledge on SE-based optical characterization of 2D materials and outlines future directions spanning enhanced measurement techniques, integration with computational photonics, and the adoption of artificial intelligence to accelerate ellipsometric data interpretation. 
\keywords{Spectroscopic Ellipsometry; Two-Dimensional Materials; Transition-Metal Dichalcogenides; Optical Constants; Excitons; Anisotropy; Hyperbolic Materials.}
\end{abstract}

\section{Introduction}
The advent of low-dimensional materials—most notably graphene and Transition Metal Dichalcogenides (TMDs) \cite{Munkhbat2022}—has catalyzed a rapid expansion of research across physics, materials science, and nanotechnology \cite{Yoo2022}. Two-dimensional (2D) materials, often consisting of a single atomic layer in their monolayer form, exhibit emergent properties that differ fundamentally from those of their bulk counterparts. For instance, TMDs such as MoS$_2$ and WS$_2$ undergo a transition from an indirect band-gap semiconductor in the bulk phase to a direct band-gap semiconductor when reduced to a monolayer \cite{Nguyen2024,Li2014, Mak2010,Ermolaev2020}. 
This remarkable evolution in their electronic structure, combined with exceptional optical and transport characteristics, positions 2D materials as leading candidates for next-generation optoelectronic and photonic devices \cite{Li2014}. Realizing this technological potential, however, necessitates a deep understanding of their intrinsic optical response—particularly the complex dielectric function, $\epsilon(\omega)$, which serves as a fundamental descriptor of their electronic and excitonic structure \cite{Yoo2022}.

Despite their immense promise, accurately characterizing 2D materials remains challenging. Their excitonic landscape—including neutral excitons, trions (charged excitons), and higher-order excitonic complexes—is highly sensitive to external perturbations. Variables such as synthesis technique (e.g., MOCVD, APCVD, LPCVD), substrate choice (e.g., Si/SiO$_2$, quartz, nickel), dielectric screening, and temperature can significantly alter their optical spectra, often complicating peak assignment and comparative analysis \cite{Nguyen2024,Maulina2024}.
Furthermore, the atomically thin nature of these materials results in an exceedingly short optical interaction path length, causing conventional optical techniques to suffer from limited sensitivity and reduced signal-to-noise ratios \cite{Yoo2022,Maulina2024}. These limitations underscore the need for characterization methods capable of resolving subtle optical features with high accuracy.

Spectroscopic Ellipsometry (SE) has emerged as one of the most powerful techniques to address these challenges. As a non-invasive, non-destructive, and highly sensitive optical method, SE enables precise characterization of surfaces, ultrathin films, and interfacial structures \cite{Politano2023,Aulika2025}. Unlike traditional reflectance or transmittance measurements, SE monitors changes in the polarization state of light upon reflection or transmission, allowing direct extraction of optical constants—namely, the refractive index ($n$) and extinction coefficient ($k$)—as well as film thickness with sub-nanometer precision \cite{Politano2023}.
Its extreme sensitivity to anisotropy, interlayer coupling, and interface quality makes SE particularly well-suited for probing the subtle optical responses of 2D semiconductors, metallic layers, and van der Waals heterostructures.

Over the past decade, the use of SE for studying 2D materials has expanded significantly, encompassing graphene \cite{Maulina2024}, monolayer TMDs \cite{Nguyen2024,Li2014}, multilayer and strongly anisotropic TMD structures \cite{Munkhbat2022}, and even emerging organic or hybrid films \cite{Aulika2025}. These studies have uncovered complex optical phenomena, including extreme anisotropy, intrinsic hyperbolicity, and the resolvable splitting of exciton–trion features.
Despite this progress, the literature remains fragmented. Many studies focus narrowly on specific materials, fabrication strategies, or modeling approaches, resulting in a lack of integrated understanding across the broader field \cite{Munkhbat2022}. A comprehensive review that synthesizes experimental methodologies, optical modeling techniques, and consolidated dielectric-function datasets across major 2D material systems is still limited.
Accordingly, the objective of this review is to provide a systematic and holistic overview of the application of Spectroscopic Ellipsometry for elucidating the optical properties of 2D materials. Specifically, this review aims to synthesize the experimental and analytical methodologies employed in SE studies of 2D systems; integrate key findings regarding optical behavior—including the dielectric function, anisotropy, excitonic resonances, and substrate effects—across representative classes of 2D materials such as graphene and both monolayer and multilayer TMDs; and discuss the remaining challenges, methodological limitations, and future research directions that may shape the advancement of SE-based characterization of 2D materials.

\section{Methodology}

\subsection{Literature Review Methodology}
This review was conducted through a systematic examination of primary scientific literature, focusing on peer-reviewed research articles published between 2014 and 2025. Comprehensive searches were performed across major databases using keywords including \textit{"Spectroscopic Ellipsometry"}, \textit{"two-dimensional materials"}, \textit{"TMDs"}, \textit{"graphene"}, \textit{"optical properties"}, and \textit{"dielectric constants"}. From this corpus, ten seminal and highly representative articles were selected to form the analytical backbone of this synthesis. These articles were chosen based on their direct relevance to SE-based characterization of 2D materials and their coverage of diverse material classes (graphene, monolayer and multilayer TMDs, organic films), methodologies (from fundamental modeling to advanced analysis), and key optical phenomena. The selected references, cited throughout this work, are \cite{Aulika2025, Li2014, Maulina2024, Munkhbat2022, Nguyen2024, Politano2023, Yoo2022}.

\subsection{Fundamental Principles of Spectroscopic Ellipsometry}
\begin{figure}[h!]
    \centering
    \includegraphics[width=0.7\linewidth]{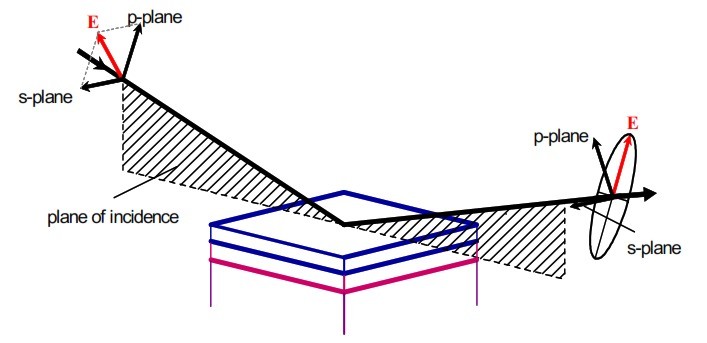} % Assuming the figure file is named f1.jpg
    \caption{Schematic of SE measurement principle showing the transformation of polarized light upon sample reflection.}
    \label{fig:se_principle}
\end{figure}

Spectroscopic Ellipsometry is a powerful, non-contact optical technique that does not measure light intensity directly. Instead, it quantifies the change in the polarization state of light after its interaction with a sample \cite{Politano2023, Yoo2022}. The technique is predicated on measuring two fundamental parameters: $\Psi$ (psi) and $\Delta$ (delta). The parameter $\Psi$ describes the amplitude ratio change between the p-polarized (parallel to the plane of incidence) and s-polarized (perpendicular to the plane of incidence) components of the reflected or transmitted light. The parameter $\Delta$ describes the corresponding phase difference introduced between these two components. These quantities are combined into the complex ellipsometric ratio $\rho$, defined as:
\begin{equation}
    \rho = \frac{r_p}{r_s} = \tan(\Psi)e^{i\Delta},
\end{equation}
where $r_p$ and $r_s$ are the complex Fresnel reflection coefficients for p- and s-polarized light, respectively \cite{Politano2023, Yoo2022}. The measurement is typically performed at oblique angles of incidence (commonly between $45^\circ$ and $80^\circ$) to maximize sensitivity to thin-film properties \cite{Aulika2025, Politano2023}. A schematic of the SE measurement principle is shown in Figure \ref{fig:se_principle}.

\subsection{Sample Preparation and Data Acquisition}
The optical properties extracted from 2D materials are critically dependent on sample quality and preparation. Two primary fabrication routes dominate: mechanical exfoliation and chemical vapor deposition (CVD). Mechanical exfoliation, often using a polydimethylsiloxane (PDMS) stamp for dry transfer, yields high-quality, pristine flakes ideal for fundamental studies of monolayers and few-layers \cite{Li2014, Munkhbat2022, Funke2016}. In contrast, CVD enables the growth of large-area, uniform films with controlled thickness, such as multilayer graphene on nickel \cite{Maulina2024} or monolayer TMDs on Si/SiO$_2$ substrates \cite{Nguyen2024}. Substrate choice (e.g., Si/SiO$_2$, quartz, ITO) is a crucial variable, as its optical properties significantly influence the measurement \cite{Aulika2025, Li2014, Munkhbat2022}.

SE data ($\Psi$, $\Delta$) were acquired using advanced commercial instruments like dual-rotating compensator ellipsometers or variable-angle spectroscopic ellipsometers (VASE) \cite{Aulika2025, Maulina2024, Munkhbat2022, Nguyen2024}. Measurements were typically conducted over a broad spectral range from ultraviolet to near-infrared (e.g., 210–1700 nm) and at multiple angles of incidence (e.g., $60^\circ$, $65^\circ$, $70^\circ$, $75^\circ$) to enhance data reliability and constrain model parameters \cite{Aulika2025, Maulina2024, Politano2023}. Complementary techniques including Raman spectroscopy, photoluminescence (PL), and scanning electron microscopy with energy-dispersive X-ray spectroscopy (SEM-EDX) were routinely employed to corroborate layer number, crystal quality, and sample morphology, thereby validating the SE analysis \cite{Li2014, Maulina2024, Nguyen2024}.

\subsection{Optical Modeling and Data Analysis}
Extracting optical constants from raw SE data is an inverse problem solved through iterative, model-dependent fitting \cite{Politano2023}. Recent advancements in artificial intelligence, particularly deep learning, offer promising avenues to accelerate this inverse problem, enabling rapid, high-throughput analysis of SE data \cite{Li2021}. The core process involves constructing a stratified optical model that represents the physical sample structure. Model parameters (e.g., layer thickness, dielectric function) are adjusted until the calculated ellipsometric response ($\Psi_{\mathrm{calc}}$, $\Delta_{\mathrm{calc}}$) matches the experimental data ($\Psi_{\mathrm{exp}}$, $\Delta_{\mathrm{exp}}$), with the fit quality quantified by a merit function like the mean squared error (MSE) \cite{Aulika2025, Maulina2024, Politano2023}.

\subsubsection{Modeling the Physical Structure}
\begin{figure}[h!]
    \centering
    \includegraphics[width=0.75\linewidth]{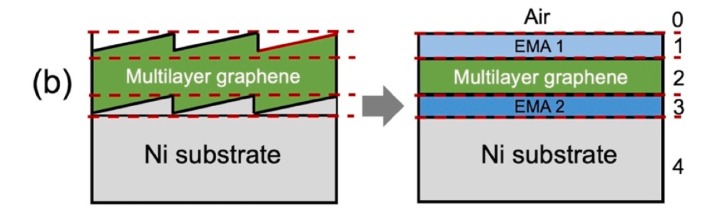}
    \caption{Mapping of the physical MLG/Ni structure (left) to a multilayer optical model (right) for SE analysis.}
    \label{fig:mlg_model}
\end{figure}

The first step is to define the physical layer stack. For 2D materials, two conceptual models are used. The \textit{three-dimensional slab model} treats the material as a homogeneous thin film with finite thickness $d$ and a bulk-like dielectric function $\epsilon$ \cite{Li2014}. The \textit{two-dimensional sheet model} represents it as an infinitely thin layer characterized by its sheet conductivity $\sigma$, circumventing thickness assignment \cite{Yoo2022, Majerus2018}. A practical example is modeling CVD-grown multilayer graphene (MLG) on Ni, which employs a stack: air / EMA surface roughness / MLG film / EMA interface / Ni substrate \cite{Maulina2024}. This mapping from physical structure to optical model is illustrated in Figure \ref{fig:mlg_model}.

\subsubsection{Accounting for Non-Idealities}
Real samples exhibit imperfections that must be incorporated into the optical model. Surface and interfacial roughness are commonly modeled using the Effective Medium Approximation (EMA), typically the Bruggeman formalism, which mixes the material and void fractions \cite{Aulika2025, Maulina2024}. Optical anisotropy is another critical consideration. Many TMDs are uniaxial, requiring a dielectric tensor with distinct in-plane ($\epsilon_{\parallel}$) and out-of-plane ($\epsilon_{\perp}$) components \cite{Munkhbat2022}. Materials with lower symmetry, like ReS$_2$, exhibit biaxial anisotropy ($\epsilon_{xx} \neq \epsilon_{yy} \neq \epsilon_{zz}$) \cite{Munkhbat2022}. Furthermore, some organic films show depth-dependent refractive index gradients, necessitating graded-index layers in the model \cite{Aulika2025}.

\subsubsection{Dispersion Models}
A physically appropriate dispersion model is required to describe the spectral dependence of the dielectric function $\epsilon(E) = \epsilon_1(E) + i\epsilon_2(E)$. All models must obey Kramers-Kronig causality relations \cite{Politano2023}. Common models include:
\begin{itemize}
    \item \textbf{Lorentz Oscillator:} For bound transitions like excitons.
    \begin{equation}
        \epsilon(E) = \epsilon_{\infty} + \sum_{k=1}^{N} \frac{f_k}{E_k^2 - E^2 - iE\gamma_k},
    \end{equation}
    where $E_k$, $f_k$, and $\gamma_k$ are the resonance energy, oscillator strength, and damping of the $k$-th oscillator \cite{Li2014}.
    \item \textbf{Drude Model:} For free-carrier response in metals or doped semiconductors.
    \begin{equation}
        \epsilon(E) = \epsilon_{\infty} - \frac{E_p^2}{E(E + i\gamma_D)},
    \end{equation}
    where $E_p$ is the plasma energy and $\gamma_D$ the scattering rate \cite{Munkhbat2022}.
    \item \textbf{Drude-Lorentz:} A hybrid model for systems with both free and bound contributions, used for graphene on metals \cite{Maulina2024}.
    \begin{equation}
        \epsilon(E) = \epsilon_{\infty} - \frac{E_p^2}{E(E + i\gamma_D)} + \sum_{k=1}^{N} \frac{f_k}{E_k^2 - E^2 - iE\gamma_k}.
    \end{equation}
    \item \textbf{Tauc-Lorentz:} For amorphous/polycrystalline semiconductors, incorporating the optical bandgap $E_g$ \cite{Politano2023}. The imaginary part is:
    \begin{equation}
        \epsilon_2(E) = 
        \begin{cases}
        \frac{A E_0 C (E - E_g)^2}{E[(E^2 - E_0^2)^2 + C^2 E^2]}, & E > E_g, \\
        0, & E \leq E_g,
        \end{cases}
    \end{equation}
    with $A$, $E_0$, and $C$ as amplitude, peak energy, and broadening. $\epsilon_1(E)$ is obtained via Kramers-Kronig integration.
    \item \textbf{Gaussian Oscillators:} A phenomenological alternative for complex materials like organic films \cite{Aulika2025}.
    \begin{equation}
        \epsilon_2(E) = A \exp\left[-\frac{(E - E_0)^2}{2\sigma^2}\right].
    \end{equation}
\end{itemize}

\subsubsection{Advanced Analysis Approaches}
For complex scenarios, advanced methods are employed. \textit{Point-by-point} extraction retrieves $\epsilon_1(E)$ and $\epsilon_2(E)$ at each energy without a pre-defined model, useful for tracking fine spectral changes like temperature-dependent exciton shifts \cite{Nguyen2024}. \textit{Critical-point analysis} uses the second derivative $d^2\epsilon/dE^2$ to enhance and resolve overlapping transitions, such as A and B excitons in TMDs \cite{Nguyen2024}. When both thickness $d$ and $\epsilon(E)$ are unknown, an iterative self-consistent procedure using multi-angle data is applied \cite{Maulina2024}. If SE data is unavailable, \textit{Kramers-Kronig constrained reflectance analysis} can extract dielectric functions from reflectance spectra by fitting with KK-consistent Lorentz oscillators \cite{Li2014}. Alternatively, machine learning approaches such as artificial neural networks can directly infer optical constants from raw ellipsometry data, significantly reducing analysis time and model dependency \cite{Li2021}.

\section{Results and Discussion}
\subsection{Multilayer Graphene on Metallic Substrates: A Case Study in Interfacial Physics}
\begin{figure}[h!]
    \centering
    \includegraphics[width=0.55\linewidth]{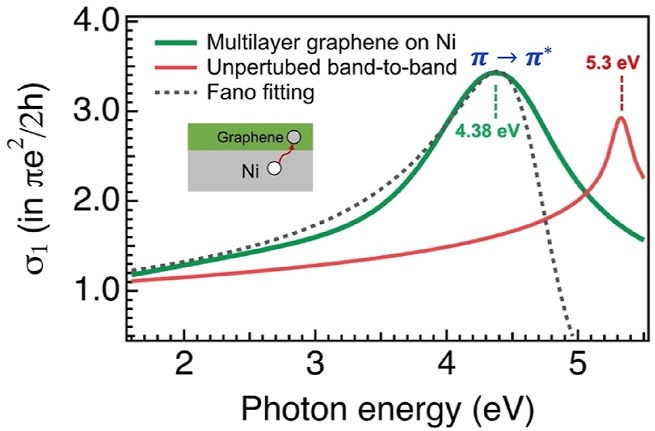}
    \caption{The real part of the optical conductivity ($\sigma_1$) for multilayer graphene on a nickel substrate. The experimental data (thick green line) shows a prominent peak at 4.38 eV, fitted with a Fano lineshape (dashed black line), representing the redshifted $\pi \rightarrow \pi^*$ transition due to strong substrate interaction. Adapted from \cite{Maulina2024}.}
    \label{fig:mlg_ni}
\end{figure}

The integration of graphene with metallic substrates is a cornerstone for many electronic and photonic applications, making the understanding of their interfacial interactions paramount. Spectroscopic ellipsometry provides a direct, non-invasive probe into these interactions by measuring the modification of graphene's intrinsic optical transitions. A seminal study on chemical vapor deposition (CVD)-grown multilayer graphene (MLG) on a nickel (Ni) substrate reveals a profound substrate effect \cite{Maulina2024}. The key spectral feature is the $\pi \rightarrow \pi^*$ interband transition, which for free-standing graphene is typically centered around 4.6 eV \cite{Kravets2010}. However, SE analysis of the MLG/Ni system identifies a distinct and symmetric peak in the real part of the optical conductivity ($\sigma_1$) at a significantly redshifted energy of 4.38 eV \cite{Maulina2024} (Figure \ref{fig:mlg_ni}). This substantial redshift of approximately 220 meV is a direct spectroscopic signature of strong graphene-substrate coupling. Broadband SE studies of CVD graphene further reveal high-energy absorption features around 6.4 eV attributed to $\sigma \rightarrow \sigma^*$ excitonic transitions \cite{Li2016}. The mechanism involves charge transfer at the interface and hybridization of graphene's $\pi$-bands with the nickel d-bands, which modifies the electronic density of states near the Fermi level and effectively lowers the energy required for the interband transition \cite{Maulina2024}. This finding is not merely a spectroscopic curiosity; it has critical implications for designing graphene-based electrodes, sensors, and heterostructures where the electronic and optical properties must be precisely tuned by the choice of underlying material. The analysis employed a Drude-Lorentz model to successfully deconvolute the contributions from free carriers (Drude) and bound interband transitions (Lorentz), showcasing the necessity of hybrid dispersion models for accurate characterization of such hybrid material systems \cite{Maulina2024}.

\subsection{Monolayer Transition Metal Dichalcogenides: The Excitonic Realm}
Monolayer TMDs (MoS$_2$, MoSe$_2$, WS$_2$, WSe$_2$) represent a paradigm shift in semiconductors, exhibiting a direct bandgap and exceptionally strong Coulomb interactions due to reduced dielectric screening \cite{Li2014, Chernikov2014}. SE has been pivotal in quantifying their optical response, which is dominated not by free-particle interband transitions but by tightly bound excitons. The dielectric function $\epsilon_2(E)$ of these monolayers is characterized by two sharp, distinct peaks labeled A and B, arising from direct gaps at the K-point of the Brillouin zone and split by valence band spin-orbit coupling \cite{Li2014, Ermolaev2020}. Remarkably, SE measurements confirm that these atomically thin layers can absorb over 15\% of incident light at the A-exciton resonance, underscoring their extraordinary oscillator strength and potential for ultra-thin photodetectors and light-emitting devices \cite{Li2014}.

\begin{figure}[h!]
    \centering
    \includegraphics[width=0.45\linewidth]{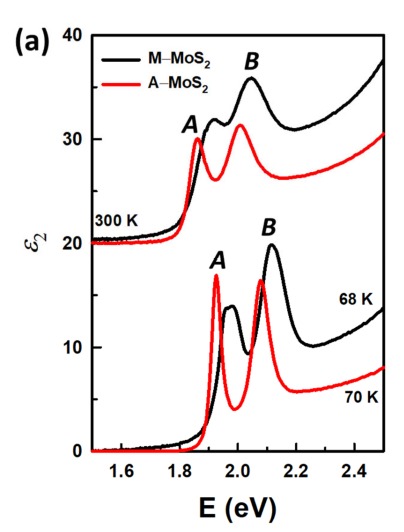}
    \caption{Imaginary part of the dielectric function ($\epsilon_2$) for monolayer MoS$_2$ at 300 K (top) and 68 K (bottom). The cryogenic spectrum reveals sharpened A and B excitons and their clear splitting into neutral (A$^0$, B$^0$) and charged (A$^-$, B$^-$) components, a feature dependent on growth method. Adapted from \cite{Nguyen2024}.}
    \label{fig:mos2_temp}
\end{figure}

The excitonic landscape is highly tunable and sensitive to both external perturbations and internal sample quality. Temperature-dependent SE studies on monolayer MoS$_2$ and WS$_2$ provide a clear picture of exciton-phonon interactions. As temperature decreases from 300 K to cryogenic levels (e.g., 68 K), the A and B exciton peaks exhibit a significant blue-shift and, more notably, a dramatic narrowing (Figure \ref{fig:mos2_temp}) \cite{Nguyen2024}. This narrowing is attributed to the "freezing out" of phonons that otherwise cause inhomogeneous broadening at room temperature. The low-temperature spectra often reveal finer structure. Crucially, SE has demonstrated that the appearance of this fine structure—specifically the splitting of the A and B resonances into neutral excitons (A$^0$, B$^0$) and negatively charged trions (A$^-$, B$^-$)—is not universal but intimately linked to the synthesis method. For instance, MoS$_2$ monolayers grown by metal-organic CVD (MOCVD) show clear low-temperature splitting into four resolvable features, while those grown by atmospheric-pressure (APCVD) or low-pressure CVD (LPCVD) do not \cite{Nguyen2024}. This implies that the MOCVD process may introduce a specific defect profile or unintentional n-type doping that stabilizes a significant trion population. This is a powerful example of how SE can serve as a diagnostic tool for material quality and doping, linking fabrication parameters directly to many-body optical phenomena that dictate device performance in LEDs, lasers, and valleytronic applications.

\subsection{Dimensionality and Interlayer Coupling: From Bulk to Monolayer TMDs}
\begin{figure}[h!]
    \centering
    \includegraphics[width=0.59\linewidth]{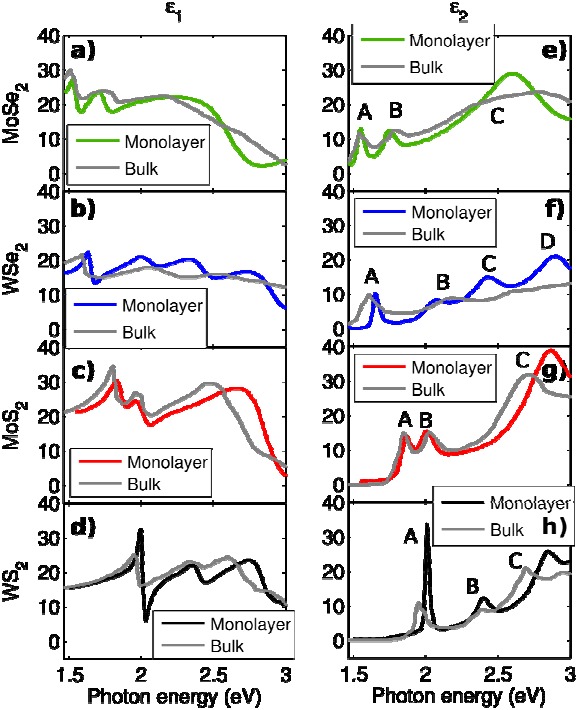}
    \caption{Comparative dielectric functions ($\epsilon_2$) of monolayer (colored lines) and bulk (gray lines) TMDs: MoSe$_2$, WSe$_2$, MoS$_2$, and WS$_2$. The plots highlight the minimal shift of the A/B excitons versus the pronounced blue-shift of the C/D transitions in monolayers, illustrating differential sensitivity to interlayer coupling. Adapted from \cite{Li2014}.}
    \label{fig:bulk_monolayer}
\end{figure}

SE enables a direct comparative analysis of how optical properties evolve with layer number, providing insights into the effects of dimensionality and interlayer coupling. Figure \ref{fig:bulk_monolayer} contrasts the dielectric functions of bulk and monolayer forms of several TMDs \cite{Li2014}. A key observation is the differential behavior of various optical transitions. The A exciton, associated with transitions largely localized on the transition metal atoms at the K-valley, shows only a modest blue-shift (10–80 meV) when going from bulk to monolayer. This relative insensitivity suggests that the electronic states involved in the A exciton are not strongly coupled between layers. In stark contrast, the higher-energy C and D transitions, which involve orbitals with greater chalcogen character and are more delocalized across the layer, undergo substantial blue-shifts of 150–300 meV in the monolayer limit \cite{Li2014}. These large shifts are direct signatures of reduced dielectric screening and quantum confinement effects that become dominant when interlayer coupling is removed. Similarly, in van der Waals heterostructures such as graphene/WS\textsubscript{2} and graphene/MoS\textsubscript{2}, SE and Raman spectroscopy reveal interfacial charge transfer and modified excitonic photoluminescence due to interlayer coupling \cite{Han2020}.. Thus, SE not only maps the optical spectrum but also provides a spectroscopic fingerprint for dimensionality, where the energy of high-energy transitions serves as a sensitive probe of layer thickness and interlayer interaction strength.

\subsection{Extreme Anisotropy and Hyperbolic Phenomena in Multilayer TMDs}
\begin{figure}[h!]
    \centering
    \includegraphics[width=0.4\linewidth]{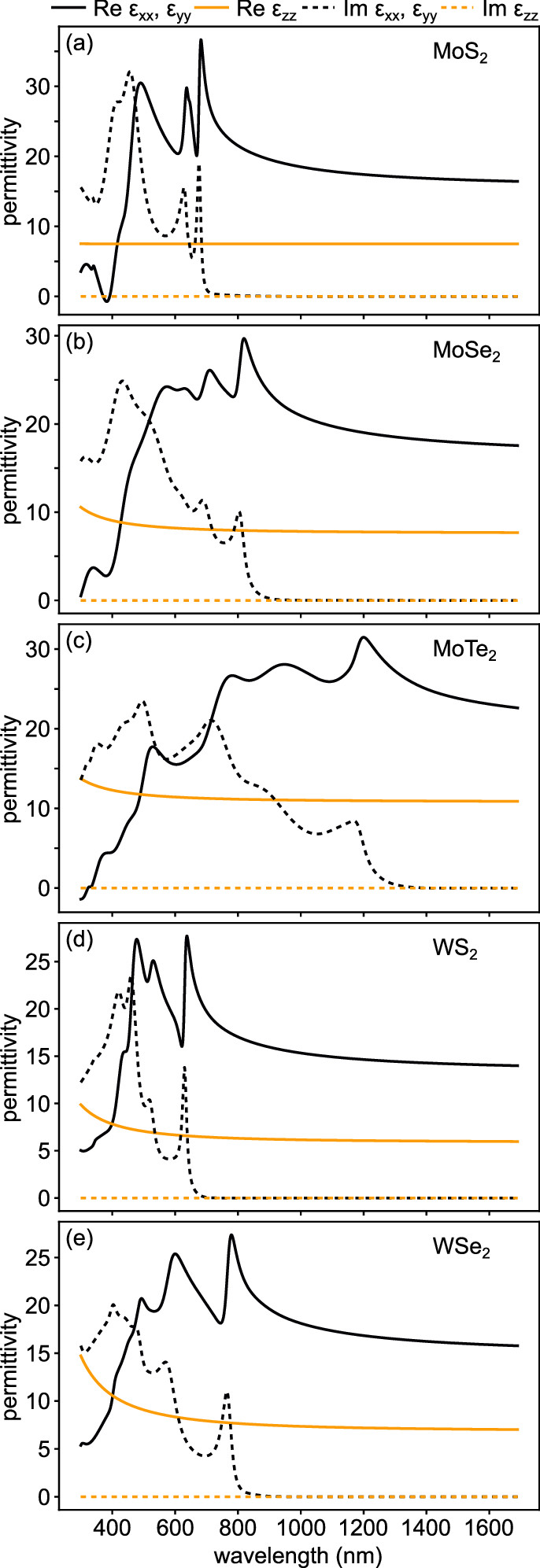}
    \caption{In-plane (thick lines) and out-of-plane (thin lines) components of the complex dielectric function for multilayer MoS$_2$, MoSe$_2$, MoTe$_2$, WS$_2$, and WSe$_2$. Solid and dashed lines represent the real ($\epsilon_1$) and imaginary ($\epsilon_2$) parts, respectively. The significant disparity between components confirms strong uniaxial optical anisotropy. Adapted from \cite{Munkhbat2022}.}
    \label{fig:anisotropy}
\end{figure}

Beyond monolayers, multilayer TMDs exhibit optical properties that are remarkable for both fundamental physics and applied photonics. SE measurements reveal that these materials possess exceptionally high in-plane refractive indices ($n_{\parallel}$) in the telecommunications-relevant near-infrared (NIR) region. For example, MoTe$_2$ reaches $n_{\parallel} \approx 4.84$ at 1550 nm, vastly exceeding traditional dielectrics like silicon ($\sim$3.47) and making them ideal for ultra-compact waveguides and high-index contrast photonic structures \cite{Munkhbat2022}. Furthermore, their layered van der Waals structure inherently leads to strong uniaxial optical anisotropy, with birefringence ($\Delta n = n_{\parallel} - n_{\perp}$) values as high as 1.54 \cite{Munkhbat2022}. This anisotropy is vividly captured in SE data, which separately resolves the in-plane and out-of-plane components of the dielectric tensor, $\epsilon_{\parallel}$ and $\epsilon_{\perp}$, as shown in Figure \ref{fig:anisotropy}.

Advanced Mueller Matrix Ellipsometry (MME), an extension of standard SE, has been crucial for studying materials with lower symmetry. In biaxial crystals like ReS$_2$, MME can disentangle the three independent dielectric tensor components ($\epsilon_{xx} \neq \epsilon_{yy} \neq \epsilon_{zz}$), revealing complex in-plane anisotropic responses \cite{Munkhbat2022}. Perhaps the most exotic discovery enabled by MME is the natural hyperbolic dispersion in certain metallic TMDs, such as TaS$_2$ and TaSe$_2$. In these materials, below a critical plasma wavelength (($\sim 1110$--$1217\,\mathrm{nm}$)), the real part of the in-plane dielectric function becomes negative ($\Re e(\epsilon_{\parallel}) < 0$) while the out-of-plane component remains positive ($\Re e(\epsilon_{\perp}) > 0$) \cite{Munkhbat2022}. This sign reversal defines a type-II hyperbolic medium, which does not require artificial nanofabrication like metal-dielectric metamaterials. Such natural hyperbolic materials support high-wavevector propagating modes (high-k modes), enabling applications in hyperlensing for sub-diffraction imaging, spontaneous emission enhancement, and novel plasmonic waveguides \cite{Munkhbat2022}. The comprehensive dielectric data plotted in Figure \ref{fig:anisotropy} for several multilayer TMDs underscore the extreme and tunable anisotropy inherent to this material family, providing a rich library for engineered metamaterials.

\subsection{Organic Thin Films: The Role of Processing and Gradients}
The principles and power of SE extend beyond inorganic 2D crystals to organic thin films, which are vital for optoelectronic devices like OLEDs and solar cells. Studies on films such as spiro-OMeTAD highlight that their optical properties are not uniform but can exhibit depth-dependent gradients in the refractive index \cite{Aulika2025}. Accurately characterizing such films requires moving beyond simple homogeneous layer models to implementing graded-index models within the SE analysis framework. Furthermore, substrate preparation, such as UV-ozone treatment, can drastically alter film formation, leading to changes in molecular ordering, density, and consequently, the degree of optical anisotropy \cite{Aulika2025}. Beyond organic films, SE is equally powerful for characterizing inorganic thin films; for instance, studies on nanocrystalline Dy\textsubscript{2}O\textsubscript{3} layers reveal thickness-dependent refractive indices and extinction coefficients critical for anti-reflective coating applications \cite{Alresheedi2021}. These findings emphasize a universal theme across all thin-film materials: the final optical response is a product of not only the intrinsic material properties but also the deposition process, interfacial chemistry, and post-processing treatments. SE serves as the critical metrology tool to quantify these effects and guide process optimization.

\subsection{Methodological Challenges and Reflections on SE for 2D Materials}
While the results presented are compelling, a critical discussion must address the challenges and limitations of applying SE to 2D materials. The technique is fundamentally an inverse problem; the raw data ($\Psi$, $\Delta$) are inverted through a physically parameterized model to obtain optical constants. This makes the analysis strongly model-dependent \cite{Politano2023, Yoo2022}. The accuracy of the extracted dielectric function is contingent upon the correct choice of the layer stack (e.g., including appropriate interface layers) and the mathematical form of the dispersion model (e.g., Lorentz vs. Tauc-Lorentz). Incorrect assumptions can lead to significant artifacts or unphysical results. For samples on transparent substrates like thick SiO$_2$/Si, optical interference within the substrate can couple the measured $\Psi$ and $\Delta$ in complex ways, making it challenging to isolate the purely intrinsic response of the ultrathin 2D layer on top \cite{Yoo2022}. Advanced analysis, such as using multiple angles of incidence, is often required to decouple these effects. Moreover, broadband SE studies have provided comprehensive dielectric functions for monolayer and bulk MoS\textsubscript{2}, revealing high permittivity and negligible absorption in the infrared range, critical for photonic applications \cite{Ermolaev2020}.

Spatial resolution is another practical limitation. Conventional SE spot sizes are on the order of hundreds of micrometers, which is suitable for CVD-grown films but too large to characterize small, exfoliated flakes or to map spatial heterogeneity within a sample \cite{Munkhbat2022, Politano2023}. Micro-spot and imaging ellipsometers address this but often with trade-offs in spectral range or speed. Finally, real-world sample imperfections—surface roughness, adsorbates, thickness variations, and inhomogeneous strain—present significant hurdles. These non-idealities must be incorporated into the optical model (e.g., using Effective Medium Approximations for roughness) to avoid misinterpreting their effects as intrinsic material properties \cite{Aulika2025}. This increases model complexity and the risk of parameter correlation, where different combinations of model parameters can yield equally good fits to the data. Therefore, SE analysis of 2D materials is most powerful and reliable when complemented by other characterization techniques like Raman spectroscopy, atomic force microscopy, and photoluminescence, which provide orthogonal information to constrain the optical model \cite{Li2014, Maulina2024, Nguyen2024}.

\section{Conclusions}
This comprehensive review has systematically examined the transformative role of Spectroscopic Ellipsometry (SE) in elucidating the fundamental optical properties of two-dimensional (2D) materials. As a non-invasive, highly sensitive, and versatile characterization technique, SE has emerged as an indispensable tool for probing the intricate electronic and photonic behavior of atomically thin systems. Through a synthesis of findings from key literature, several overarching conclusions can be drawn. Firstly, SE provides unparalleled quantitative access to the complex dielectric function, $\epsilon(\omega) = \epsilon_1(\omega) + i\epsilon_2(\omega)$, which serves as the foundational descriptor for understanding light-matter interactions in 2D materials. The technique's sensitivity to minute changes in polarization enables the precise extraction of optical constants—the refractive index ($n$) and extinction coefficient ($k$)—as well as film thickness with sub-nanometer accuracy, even for monolayers. This capability is paramount for materials like monolayer transition-metal dichalcogenides (TMDs), where SE measurements have directly quantified exceptionally strong excitonic resonances with absorption exceeding 15\% and resolved the fine spectral splitting between neutral excitons (A$^0$, B$^0$) and charged trions (A$^-$, B$^-$) under controlled synthesis and cryogenic conditions \cite{Li2014, Nguyen2024}.

Secondly, the application of SE has been instrumental in revealing how external factors and material dimensionality dictate optical responses. The review highlights the profound influence of the substrate, as exemplified by the significant redshift of the $\pi \rightarrow \pi^*$ transition in multilayer graphene on nickel due to charge transfer and interfacial hybridization \cite{Maulina2024}. It further underscores the critical role of fabrication methodology; for instance, metal-organic chemical vapor deposition (MOCVD) induces distinct trion populations in MoS$_2$ not observed in samples grown by other CVD methods, demonstrating that synthesis is a powerful knob for tuning many-body physics \cite{Nguyen2024}. The evolution from bulk to monolayer TMDs, characterized by SE, vividly illustrates the effects of quantum confinement and reduced dielectric screening, where high-energy optical transitions involving chalcogen orbitals undergo dramatic blue-shifts, while the more localized A exciton remains relatively stable \cite{Li2014}.

Thirdly, advanced SE modalities, particularly Mueller Matrix Ellipsometry (MME), have unlocked the study of complex anisotropic and hyperbolic phenomena. The review consolidates findings on the extreme optical properties of multilayer TMDs, including record-high in-plane refractive indices (e.g., $n_{\parallel} \approx 4.84$ for MoTe$_2$) and strong uniaxial birefringence, which position these materials as superior candidates for integrated nanophotonics \cite{Munkhbat2022}. Perhaps more strikingly, MME has identified naturally occurring type-II hyperbolic dispersion in metallic TMDs like TaS$_2$, where the sign reversal between in-plane and out-of-plane permittivity components enables exotic waveguiding and light confinement effects previously unattainable with conventional materials \cite{Munkhbat2022}. These discoveries expand the material library for metamaterials and transformative optical devices.

However, the journey of SE with 2D materials is accompanied by significant methodological challenges that define the current frontiers of the technique. The inherent model-dependency of SE analysis remains a primary constraint, as the fidelity of extracted optical constants is irrevocably tied to the physical accuracy of the chosen multilayer stack and dispersion model (e.g., Lorentz, Tauc-Lorentz, Drude) \cite{Politano2023, Aulika2025, Majerus2018}. This challenge is exacerbated when characterizing non-ideal samples exhibiting surface roughness, thickness gradients, or complex heterostructures, where model ambiguity can lead to substantial interpretation errors \cite{Aulika2025, Yoo2022}. Furthermore, the limited spatial resolution of conventional ellipsometers restricts their application to large-area films, leaving the characterization of small exfoliated flakes or spatially inhomogeneous samples to specialized micro-spot systems \cite{Munkhbat2022, Politano2023}. The analysis of samples on thick, transparent substrates also introduces complexities due to interference effects that couple the dielectric response components, complicating the deconvolution of the intrinsic 2D layer's properties \cite{Yoo2022}.

Looking forward, the future of SE in 2D materials research is poised at the confluence of technological innovation and interdisciplinary integration. The growing database of anisotropic optical constants for diverse 2D materials provides critical inputs for computational photonics, enabling the rational design of devices like waveguides, resonators, and metasurfaces through finite-difference time-domain (FDTD) and finite-element method (FEM) simulations \cite{Munkhbat2022, Yoo2022}. Methodologically, advancements are anticipated in in-situ and real-time SE monitoring of growth and processing, which will link optical properties directly to fabrication dynamics \cite{Politano2023}. The development of deterministic ellipsometry approaches promises more robust analysis of complex heterostructures \cite{Yoo2022}. Most transformatively, the integration of artificial intelligence and machine learning with SE data interpretation holds the potential to revolutionize the field \cite{Li2021}. These tools promise to automate model selection, recognize subtle spectral patterns indicative of specific defects or phases, and significantly reduce the reliance on expert-driven manual fitting, thereby accelerating the characterization pipeline and enabling high-throughput screening of novel 2D materials \cite{Politano2023}. In conclusion, Spectroscopic Ellipsometry has not only provided a deep window into the captivating optical world of two-dimensional materials but also continues to evolve, promising to remain a cornerstone analytical technique that drives the discovery and application of next-generation atomic-scale optoelectronic systems.

\bibliographystyle{unsrt}  


\end{document}